\documentclass[%
reprint,
superscriptaddress,
showpacs,
showkeys,
 amsmath,amssymb,
 aps,
pra,`
]{revtex4-1}

\usepackage{graphicx}
\usepackage{dcolumn}
\usepackage{bm}
\usepackage{hyperref}

\usepackage{subfigure}

\usepackage{siunitx}

\graphicspath{{Figures/}} 

\begin{document}

\title{Design of a millimetre-scale magnetic surface trap for cold atoms}
\date{\today}

\author{Dimitris Trypogeorgos}
\email[E-mail: ]{d.trypogeorgos@physics.ox.ac.uk}
\affiliation{Clarendon Laboratory, Department of Physics, University of Oxford,\\Parks Road, Oxford, OX1 3PU, UK}
\author{Stephen D. Albright}
\affiliation{Department of Physics, Brown University, Providence, Rhode Island 02912, USA}
\author{Daniel Beesley}
\author{Christopher J. Foot}
\affiliation{Clarendon Laboratory, Department of Physics, University of Oxford,\\Parks Road, Oxford, OX1 3PU, UK}

\begin{abstract}
We study a novel millimetre-scale magnetic trap for ultracold atoms, in which the current carrying conductors can be situated outside the vacuum region, a few mm away from the atoms. This design generates a magnetic field gradient in excess of \SI{1000}{G/cm} at a distance of \SI{2}{mm} from the conductors. We perform electromagnetic and thermo-mechanical characterisation using Finite Element Methods (FEM). The predicted behaviour has been verified by electrical and thermal measurements on a prototype, but has not been implemented on an apparatus with cold atoms. Operating this trap at the highest gradient allows for rapid evaporative cooling comparable to that achieved by atom chips.
\end{abstract}



\maketitle

\section{Introduction}

The confinement of atoms by magnetic fields is widely used to hold clouds of atoms in a region of high vacuum where they are well insulated from the environment. This enables atoms to be cooled by evaporation, and if the density is sufficiently high, quantum degeneracy can be achieved in dilute atomic vapour. Bose-Einstein condensation was first observed using a Time-Orbiting Potential (TOP) trap \cite{anderson_observation_1995}, and soon after in a Ioffe-Pritchard with static fields \cite{mewes_bose-einstein_1996,davis_bose-einstein_1995}. There are numerous other trapping configurations including a combination of magnetic fields and repulsive dipole forces from a laser beam (plugged quadrupole) and dressed-atom adiabatic potentials \cite{gildemeister_trapping_2010}, both of which can be used to form ring-shaped clouds as well as simply connected geometries. Instead of classifying traps by their principle of operation, they can be categorised by their method of construction. The first generation of traps were wound with water-cooled copper wire, outside the vacuum chamber in most cases, so that the current carrying conductors were at least a few centimetres from the atoms -- we call these cm-scale traps. Later, so-called atom chips were developed using microfabricated wires on the surface of a substrate inside the high vacuum only tens of micrometres away from the atoms \cite{fortagh_magnetic_2007, reichel_atomic_1999, hinds_magnetic_1999, fortagh_miniaturized_1998, schmiedmayer_quantum_1998, kruger_trapping_2003}. The advantage of having the atoms in close proximity to the wires can readily be seen by differentiating the expression obtained from the Biot-Savart law; the force on the atoms is proportional to the gradient of the magnetic field which is $B'=\left(\mu_0/2 \pi\right) I/r^2$ at a distance $r$ from a wire carrying current $I$, e.g. for a distance $r =\SI{10}{\micro m}$ and $I = \SI{1}{A}$ gives $B'=\SI{2e5}{G/cm}$. In many atom chips the conductors lie in the same plane; the use of a surface trap to give a Ioffe field was described by Weinstein and Libbrecht \cite{weinstein_microscopic_1995} which resembles the design presented here. However, shortcomings of atom chips include the restriction of the optical access when there is a surface close to the atoms. 
\begin{figure}[t]
\centering
\includegraphics[width=\columnwidth]{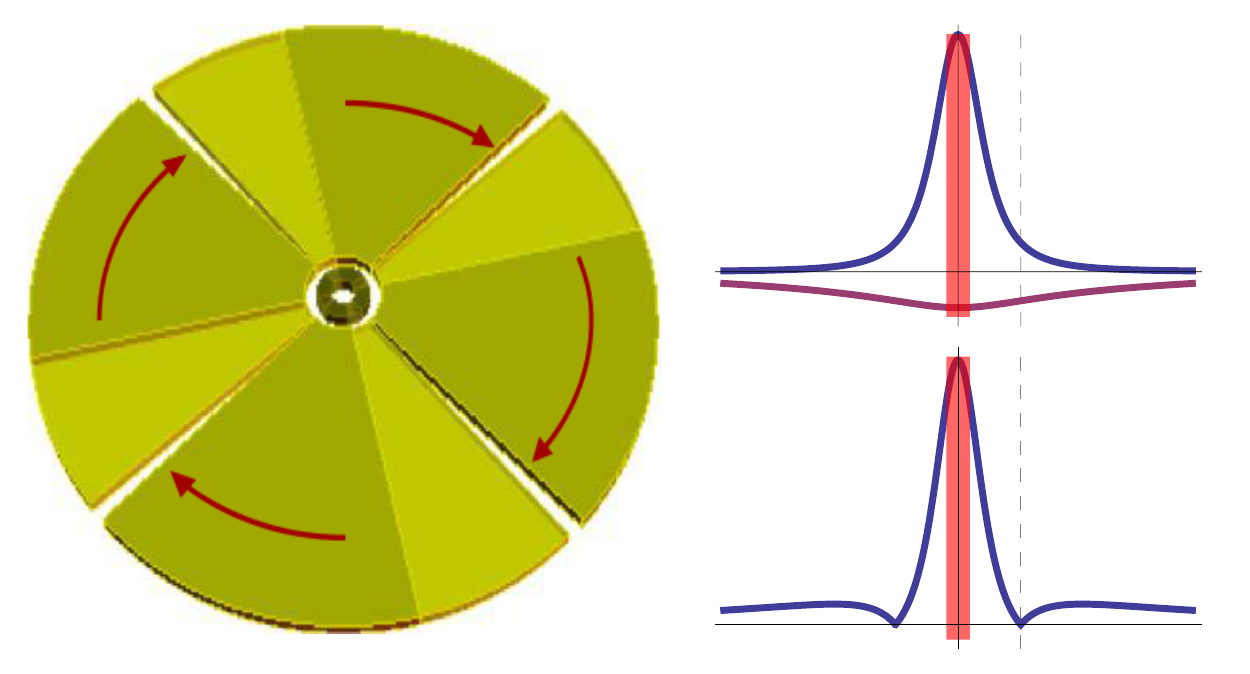}
\caption{Current flows counterclockwise in the the inner ring and in the opposite sense in the outer one. The current flows only in the segments with dark shading, i.e. there is only a fraction of the field that would be produced by the full rings. Although seemingly unphysical, this corresponds to the actual situation where the current flows are more complicated. The current density is assumed to be constant in these regions. The inner, outer radii  are \SI{1}{mm}, \SI{2}{mm}, and \SI{3}{mm}, \SI{24}{mm} for the inner and outer ring respectively. The plots on the right hand side show the field from each ring (upper) and the total field of the configuration (lower). The field is zero at a point about \SI{2}{mm} from the surface of the conductors. The extend of the \SI{2}{mm} thin conductor is shown in red block shading.}
\label{fig:coils}
\end{figure}
\begin{figure}[ht]
\centering
\subfigure[]{
\includegraphics[width=.38\columnwidth]{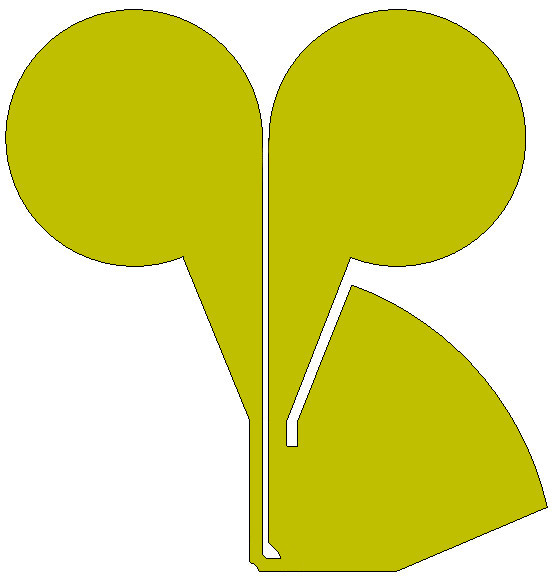}
\label{fig:trapflat}
}
\subfigure[]{
\includegraphics[width=.55\columnwidth]{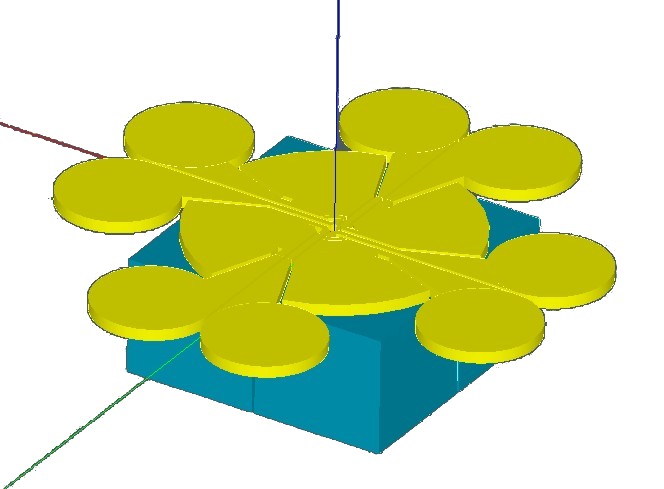}
}
\caption{Design of the cloverleaf surface trap. (a) The trap comprises of four identical slices with electrical connections made at the large disks around the outer circumference. The various cuts in the trap force the current flow to be clockwise on the outer ring and anticlockwise on the inner one as it flows between the two electrical connections. The radii of the rings are the same as in Fig.~\ref{fig:coils}. (b) A schematic of the assembled trap mounted on a cooling block. The block is divided into four electrically isolated pieces in order to avoid unwanted eddy currents.}
\label{fig:trap}
\end{figure}
Also the region of space over which the trapping field extends is relatively small and an auxiliary trap is required to load atoms into the small capture volume. A key advantage of the small scale traps is their much lower power dissipation, e.g., to reach gradients of \SI{1000}{G/cm} a typical cm-scale trap will dissipate electrical power of several kW, which although readily manageable in a laboratory environment with suitable electrical power supplies and cooling, is somewhat cumbersome.  There has been some work with conductors of intermediate size, which we call mm-scale traps, notably \cite{gupta_bose-einstein_2005, jollenbeck_hexapole-compensated_2011, hopkins_proposed_2004, wang_reaching_2007}, and we give details of a new trap of this type in this paper. The design goals were to achieve the required magnetic field gradient with current conductors outside the vacuum region, using only a moderate electrical power of around \SI{1}{kW}. We assume that the atoms are \SI{2}{mm} from the current carrying conductors, i.e. the vacuum windows are thin ($<\SI{1}{mm}$). Moreover the design allows for a \SI{1}{mm} diameter hole through which a laser beam can be directed along the symmetry axis of the trap, thus making it compatible with optical lattice experiments. In such experiments the beams forming the lattice typically have a waist size of \SIrange{100}{300}{\micro m} to facilitate overlap and give a long Rayleigh range. In this paper we report on the design and preliminary testing of the thermal loading and field measurements, but not actual trapping of atoms. However confinement of cold atoms is possible by various approaches using a magnetic quadrupole field and a suitable ultra-high vacuum apparatus.

\subsection{Principle of operation}

Quadrupole traps are often formed by using two coaxial coils of equal dimensions with current flowing in opposite directions. In contrast, this design generates a quadrupole field using two concentric rings of different radius (lying on the same plane) \cite{jian_double-loop_2013} as illustrated in Fig.~\ref{fig:coils}. The coils are connected in series so that the current flowing through them is exactly the same. We calculate the magnetic field using the Biot-Savart law. Integrating the contributions to the field over the radial and angular coordinates of a current carrying loop gives the field along the symmetry axis as follows:
\begin{equation}
B=\mu_0\sqrt{\frac{P\lambda}{r_1\rho}}G_z
\label{eq:mag}
\end{equation}\\
where $P$ is the total power dissipated in the conductor, $\lambda$ is a numerical factor which accounts for the non-uniform distribution of current in the conductor, $\rho$ is the resistivity of the conductor, $r_1$ is the inner radius of the conductor, and
\begin{widetext}
\begin{equation}
G_z=\sqrt{\frac{1}{8\pi\beta(\alpha^2-1)}}\left( (\gamma+\beta)\log\frac{\alpha+\sqrt{\alpha^2+(\gamma+\beta)^2}}{1+\sqrt{1+(\gamma+\beta)^2}}-(\gamma-\beta)\log\frac{\alpha+\sqrt{\alpha^2+(\gamma-\beta)^2}}{1+\sqrt{1+(\gamma-\beta)^2}}\right)
\end{equation}
\end{widetext}
This unitless geometric factor depends on $\alpha=r_2/r_1$ which is the ratio of the outer radius of the conductor $r_2$ to the inner radius $r_1$, $\beta=l/2r_2$ which is the ratio of the thickness of the conductor $l$ to the outer radius and $\gamma=(z_2+z_1)/2r_1$, a rescaled coordinate with $z_1$, $z_2$ being the distance to the point where the magnetic field is measured on axis from the close and the far end of the conductor respectively. In our configuration, the total field produced is simply calculated as the sum of two rings with opposite currents and different sizes.

The non-uniform current density distribution in the actual trap leads to a reduction in the magnetic field as compared to the field produced by two ring conductors with uniform current density. Fitting the data from the finite element analysis with Eq.~\ref{eq:mag} we find $\lambda=0.39$. The atoms are trapped near the zero of the magnetic field where the field contributions from the two rings with opposite currents cancel each other. We simulated the behaviour of this current distribution using the software package RADIA from the European Synchrotron Radiation Facility \cite{chubar_three-dimensional_1998}. As can be seen in Fig.~\ref{fig:coils}, both rings give magnetic fields that have a maximum at the centre but the field from the smaller one falls off more rapidly with distance than that of the larger ring. The position where the two fields cancel depends on the geometry of the conductors and the current flow.

Cold atoms in a magnetic trap follow the field adiabatically so that the tapping potential only depends on the magnitude of the field. Atoms are only trapped in weak-field seeking states beacause a static magnetic field cannot have local maxima in free space; since such states are not the minimum energy states, atoms may be ejected from the trap by inelastic collisions and can be lost from the trap through Majorana transitions when they pass near a point where the field is zero \cite{sukumar_spin-flip_1997}. One way to plug the leakage of atoms through Majorana spin flips is by creating a TOP trap \cite{anderson_observation_1995}. This can be implemented by modulating the current of each quadrant of this trap or by the addition of a rotating bias field generated using external coils.

\section{Design of the Cloverleaf Surface trap}

Based on the above considerations we designed a trap capable of producing high magnetic field gradients and be mounted outside the vacuum apparatus. In each of the four identical pieces shown in Fig.~\ref{fig:trap}, current flows in through one contact point (disk) and out through its adjacent one. We designed the trap using freely available finite element software included in the CAELinux distribution \footnote{The Computer Aided Engineering Linux distribution \href{http://www.caelinux.com/CMS/}{http://www.caelinux.com/CMS/} is based on Ubuntu and includes a plethora of open-source software. The SALOME project \href{http://www.salome-platform.org/}{http://www.salome-platform.org/} aims to develop an open-source CAD-to-FEA link based on Open CASCADE. Elmer \href{http://www.csc.fi/english/pages/elmer}{http://www.csc.fi/english/pages/elmer} is an open-source multiphysical simulation software that uses the Finite Element Method to solve the partial differential equations describing a physical system.}. Our optimised design generates a quadrupole magnetic field with a gradient in excess of \SI{1000}{G/cm} centred at a distance of \SI{2}{mm} from the top surface of the conductors.
\begin{figure}[h]
\centering
\subfigure[]{
\includegraphics[width=.46\columnwidth]{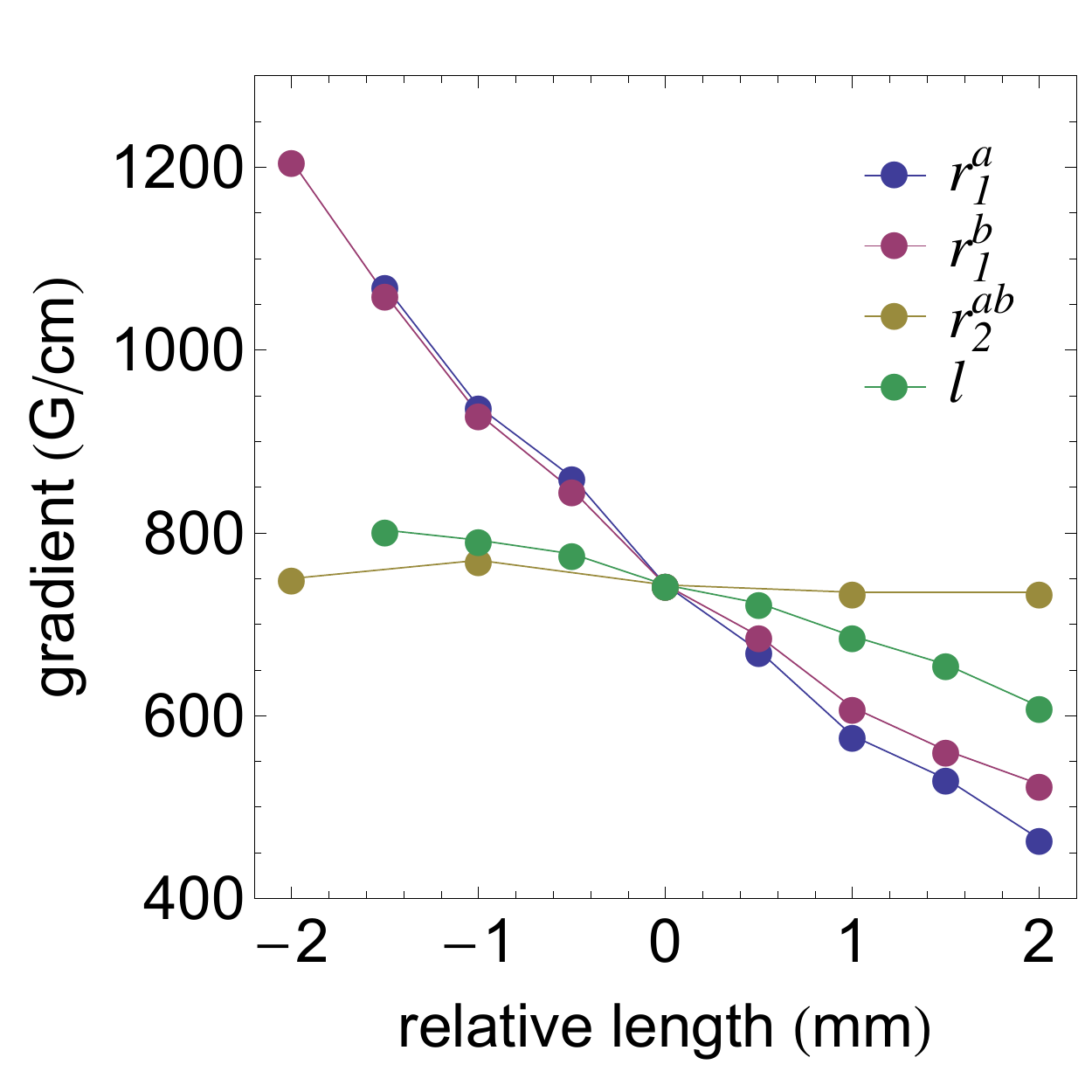}
}
\subfigure[]{
\includegraphics[width=.46\columnwidth]{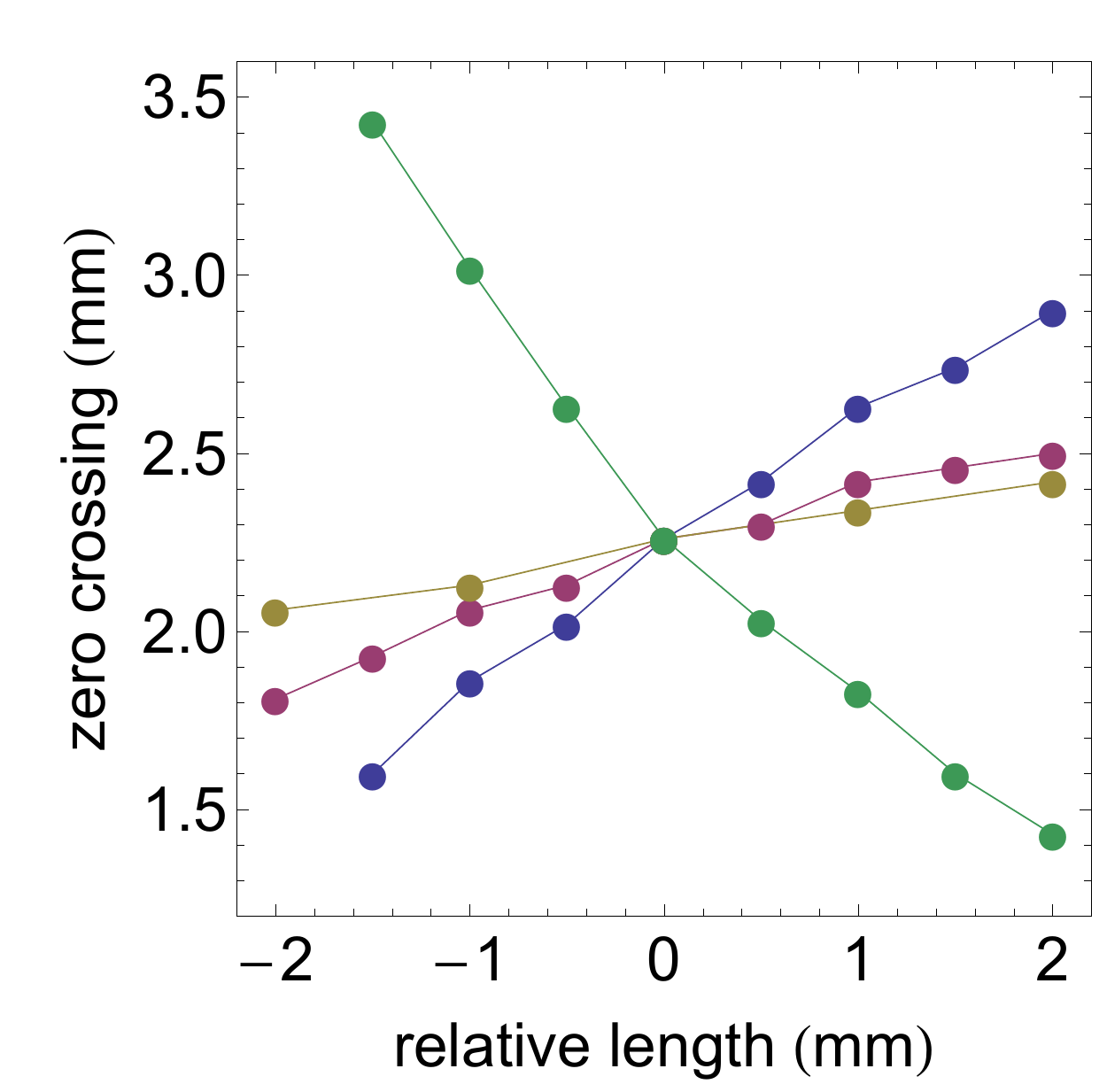}
}
\caption{Effect of varying the relevant dimensions of the trap on (a) the magnetic field gradient and (b) the point where the magnetic field is zero measured from the surface of the trap. The dimensions are referenced to the initial conditions that were chosen to be the dimensions of the two concentric rings (see Fig.~\ref{fig:coils}). }
\label{fig:dim}
\end{figure}
We used the approximation of two current-carrying rings (as in Fig.~\ref{fig:coils}) to determine starting values of the optimisation routine. Variations of the parameters relative to the initial values gave the trends shown in Fig.~\ref{fig:dim}. The position where the field is zero increases with increasing radii. Decreasing the thickness has the same effect albeit much more pronounced. The gradient increases when decreasing any of the dimensions with the thickness playing the most important role along with the dimensions of the outer ring. With these trends in mind we chose the three innermost radii to be as small as possible and scaled $r_2^b$ to increase the point where the field is zero. We also try to keep $l$ as small as realistically possible. The dimensions of the final trap are $r_1^a=\SI{1}{mm}$, $r_2^a=\SI{2}{mm}$,  $r_1^b=\SI{3}{mm}$, $r_2^b=\SI{24}{mm}$ and $l=\SI{2}{mm}$. These values optimise the gradient for a given current at a fixed distance from the surface of the conductors.
 
\begin{figure}[t]
\centering
\includegraphics[width=\columnwidth]{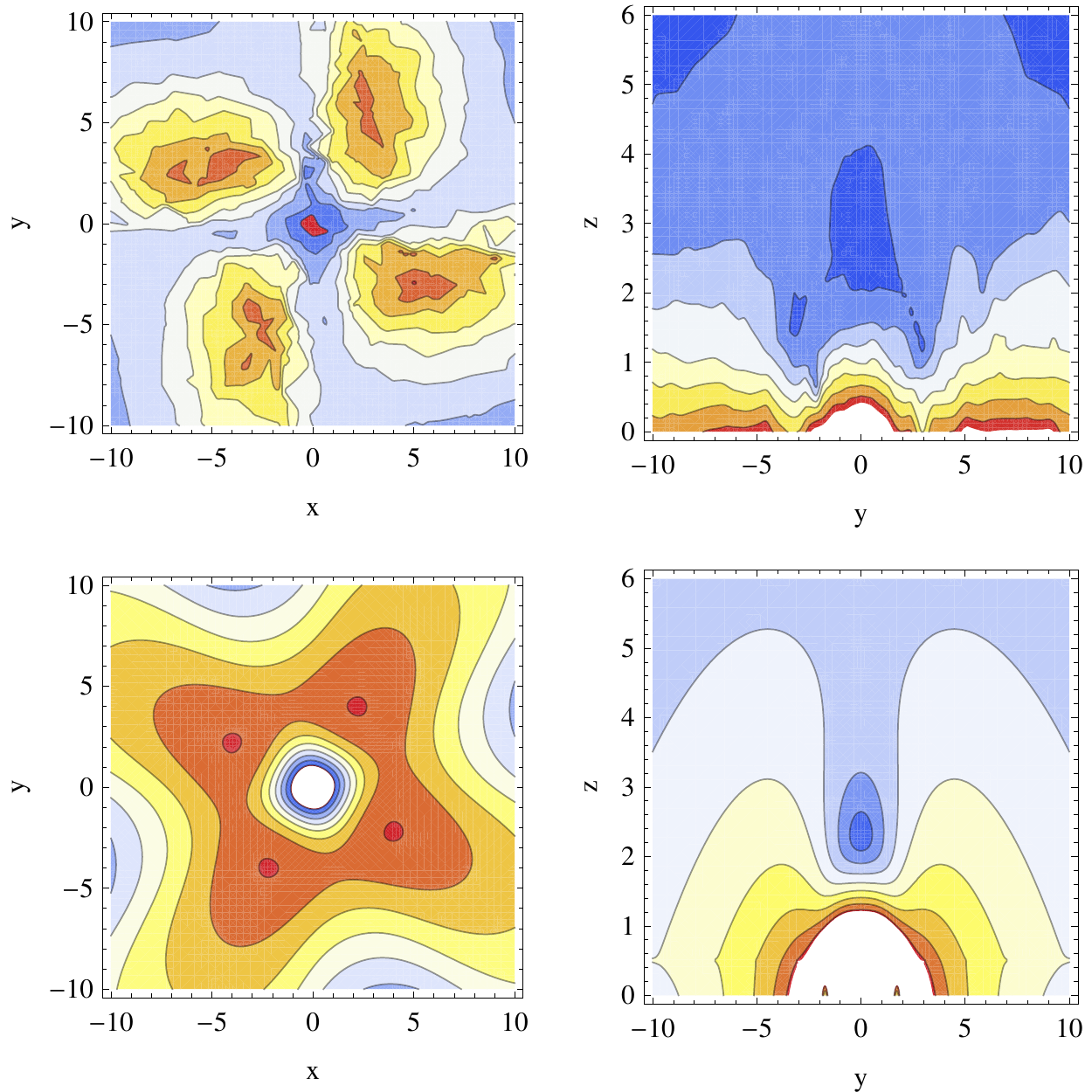}
\caption{Contour plots of the magnetic field amplitude along the xy-plane (first column) and yz-plane (second column). The top row figures show the FEM output while the bottom ones the idealised results from the Radia software. The effect of the non-uniform current density can be clearly seen on the xy-plane. The grid size used for the FEM results was chosen as fine as possible consistent with reasonable simulation times.}
\label{fig:cont}
\end{figure}

\begin{figure}[t]
\centering
\subfigure[]{
\includegraphics[width=.46\columnwidth]{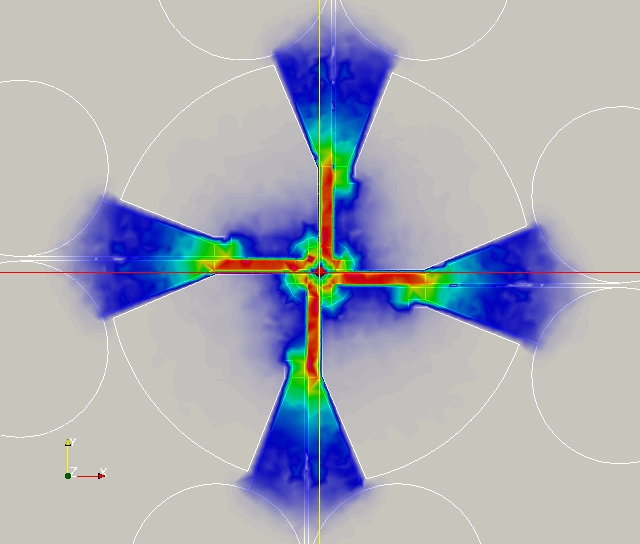}
\label{fig:currd}
}
\subfigure[]{
\includegraphics[width=.46\columnwidth]{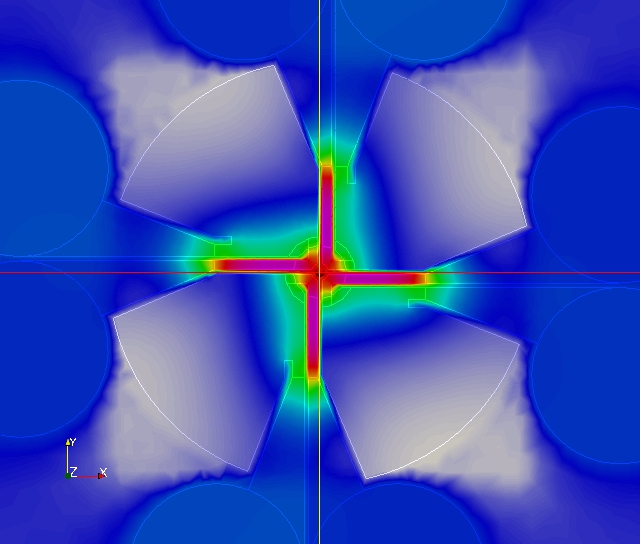}
\label{fig:temp}
}
\caption{(a) The current density distribution and (b) the temperature distribution are closely related. The maxima of both (red regions) correspond to the narrow parts of the conductors of cross section \SI{2x2}{mm}. According to the FEM simulations, the maximum temperature should not exceed \SI{50}{\celsius} with the cooling block at \SI{12}{\celsius}.}
\label{fig:currT}
\end{figure}

The magnetic field of the optimised trap is shown is Fig.~\ref{fig:cont}. The FEM shows the true distribution taking into account the steady-state non-uniform current density within the conductors, including the effect of their shape and temperature variations with position. For comparison we also plot the idealised magnetic field distribution. It is interesting to note in Fig.~\ref{fig:cont} is that there is a rotation of the field on the xy-plane. This is due to the profile of the current distribution (Fig~\ref{fig:currd}) which is dictated by the shape of the trap. The current density increases significantly at the conical entry point and that manifests as an apparent rotation of the magnetic field. 
The trap behaves like a surface trap albeit at a much larger scale. Its fourfold symmetry does not give rise to multipole orders of the field higher than the quadrupole.

\subsection{Construction of the trap}

The trap is precision machined from a \SI{2}{mm} thick copper plate in four identical slices using Wire Electrical Discharge Machining. Dielectric is placed between the cuts to keep the slices isolated. The width of the cuts has been optimised to balance the conflicting demands of low local temperature and strong, uniform and symmetric magnetic field. The conductors are fixed onto a cooling block which also serves as a mechanical mounting. The current carrying cables are attached to the trap symmetrically to minimise field distortions. Additional shim coils are used to counter external bias fields.
The trap can be cooled very efficiently using a cooling block with chilled water running through it attached on one side of the trap. The cooling block in FEM is modelled as a block of copper that has water flowing through it at a pressure of \SI{4}{bar} and temperature of \SI{12}{\celsius}. There is a large thermal conductance because of the large surface area which leads to efficient cooling. FEM show that the trap stays at less than \SI{50}{\celsius} with \SI{400}{A} running through it. More importantly, due to its small thickness, cooling should be uniform across the volume of the trap although some warmer spots are generated where the current density is high. Nevertheless, these are symmetric and therefore do not distort the field. A significant advantage over using two separate coils is that any fluctuations coming from non-laminar water flow or uneven heat extraction are common mode, which increases the stability of the trap. Numerical simulations of the magnetic field show that distortions of the field due to corrugations on the surface of the trap are negligible. The natural length scale of the trap is a few millimetres, much larger than features arising from surface roughness. 

\section{Measurements}

The magnetic field was measured using a Hall sensor (634ss2 from Honeywell in a 4-pin DIP package). This sensor was chosen due to its sensitivity and small size. Although the height of the DIP package is about \SI{2.2}{mm} we were able to measure sub-millimetre field changes with it since the active area of the sensor is much smaller and located at about \SI{1.5}{mm} from its surface.
\begin{figure}[]
\centering
\includegraphics[width=\columnwidth]{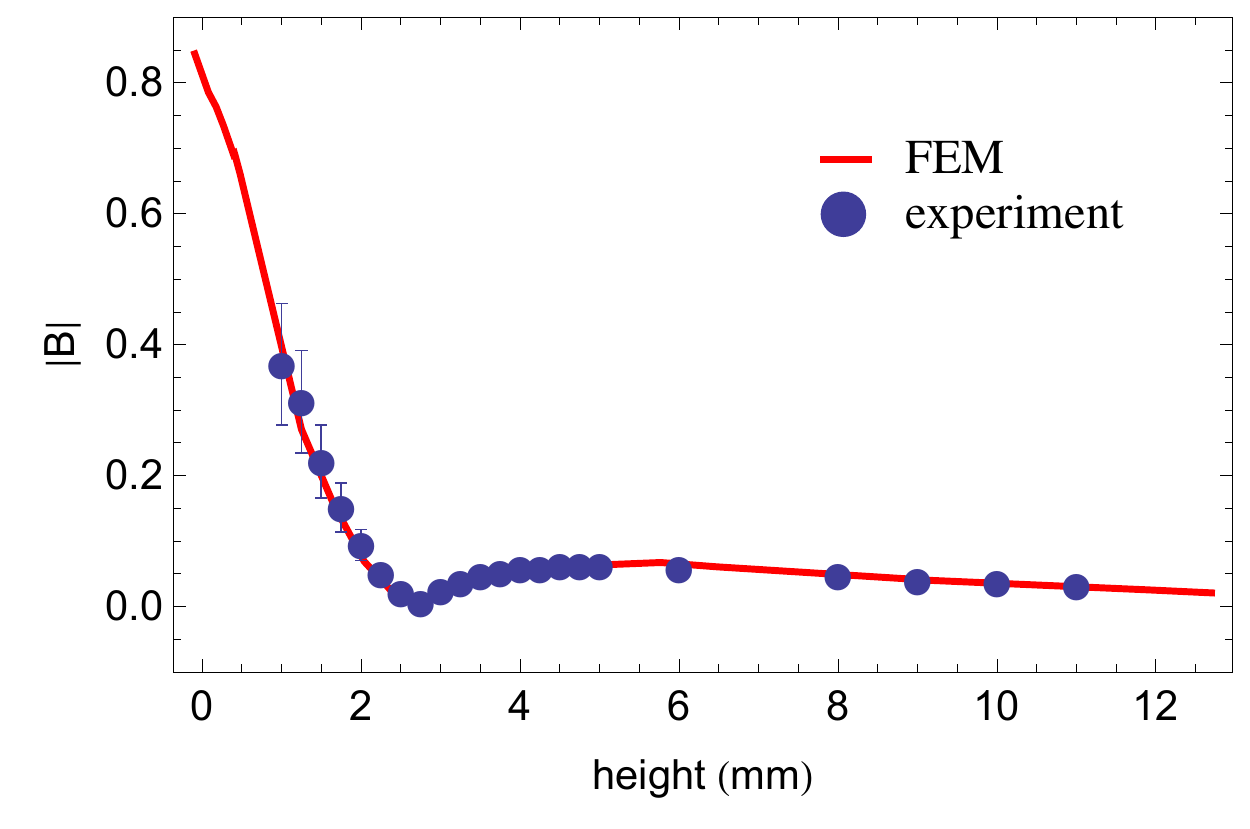}
\caption{A comparison of the measured magnetic field amplitude (data points) against the numerically calculated profile (red line). All data are normalised to the maximum magnetic field at the centre of the trap. The active area of the Hall sensor is taken to be at \SI{1.5}{mm} from its surface. For this reason no data were taken at distances smaller than \SI{1.5}{mm}. The height is measured from the centre of the trap. The agreement of the experimental data with the the theory is excellent. Uncertainties arise mainly from the probe resolution and are roughly proportional to the magnitude of the field.}
\label{fig:Bexp}
\end{figure}
The measurement of the magnetic field was done with a phase sensitive detection scheme to overcome the high frequency electronic noise ($>\SI{1}{kHz}$) and various DC offsets that were present; a current of \SI{1.8}{A} switched at \SI{20}{Hz} was used that was well within the bandwidth of the Hall probe. The output of the Hall probe was amplified by an active low pass filter with $\tau=\SI{1}{ms}$. After accounting for the tolerances of the passive elements and the losses of the phase analyser, the sensitivity of the measurement was found to be $(8 \pm 2) \times 10^5 \SI{}{mV/T}$. This was more than adequate for the strength of the fields produced at the chosen current. As shown in Fig.~\ref{fig:Bexp} the measured values agree extremely well with the expected ones.

\begin{figure}[]
\centering
\includegraphics[width=.96\columnwidth]{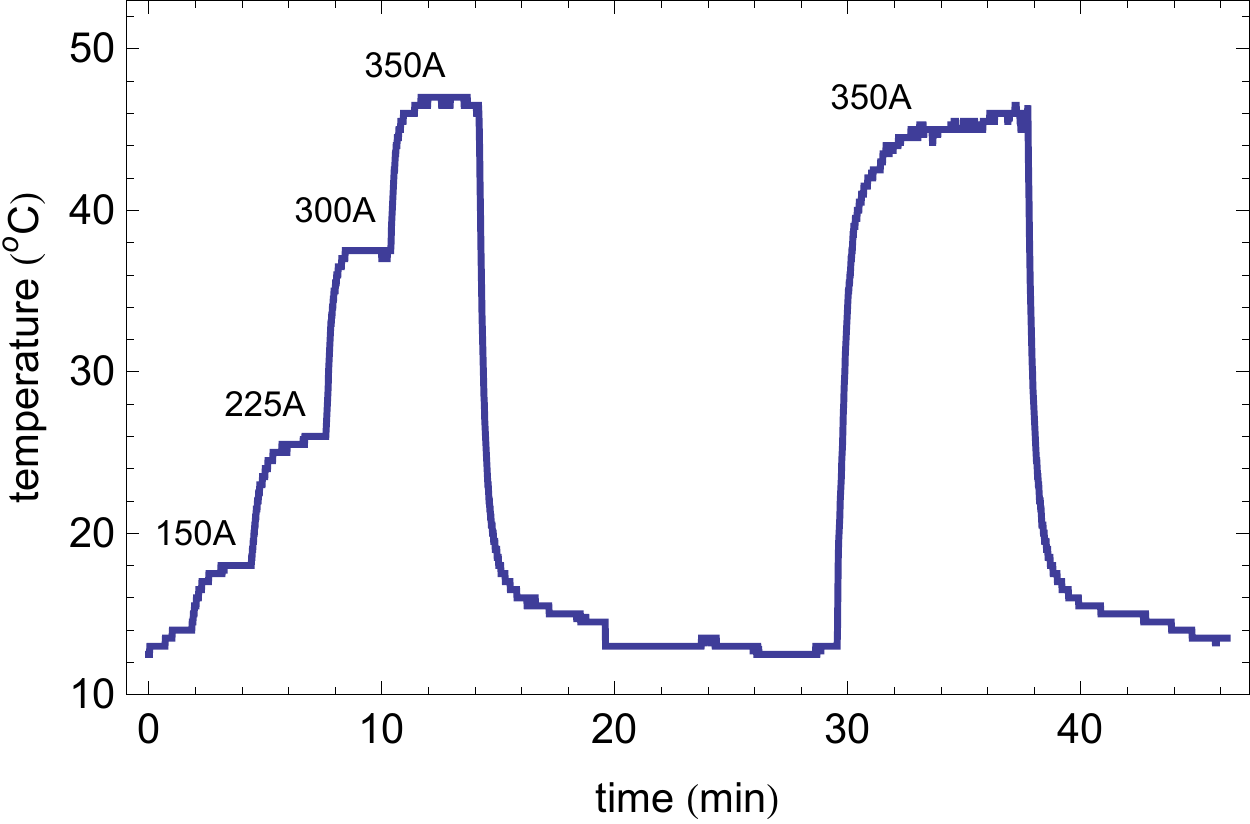}
\caption{Measurement of the temperature of the trap for various values of the applied current. The temperature is measured at the hottest point of the trap (see Fig.~\ref{fig:temp}). In the first cycle the temperature is left to stabilise for the shown current values. The second cycle represents a more realistic operation of the trap where the maximum current is applied instantly.}
\label{fig:tempAll}
\end{figure}

We measured the temperature distribution using a standard type K temperature probe. Since the thinnest parts of the trap present the highest resistance they tend to get hotter (as can be seen in Fig.~\ref{fig:temp}). The trap is water cooled directly using water at \SI{14}{\celsius} and \SI{10}{lt/min} flow (a slightly different configuration than assumed in FEM simulations). The trap was connected to the high current leads using \SI{7}{cm} long copper flats to ensure a good electrical connection and also distance it from the current leads since it was found that heat from the leads was increasing the steady state temperature of the trap. First, we increased the current in steps of \SI{75}{A} up to a maximum of \SI{350}{A} and let the temperature stabilise at each step. After switching the current off and waiting for the temperature to stabilise again the maximum current was run through the trap and the cycle repeated. One period of this cycle can be seen in Fig.~\ref{fig:tempAll}. The maximum temperature of about \SI{48}{\celsius} was reached at a lower current than expected from the numerical simulations, \SI{350}{A} cf. \SI{400}{A}. This is mainly due to the thermal conductance being less than assumed in the FEM simulations. Therefore, a system where water in direct contact with the surface of the copper conductors was used to cool the trap and achieve design specifications. We note that this is the maximum steady state temperature that is attained after about \SI{10}{min} of operation of the trap at maximum current. In practice we expect the trap to be operated only for a few minutes per experimental run so its working temperature will be even lower.

\section{Conclusion}

We have demonstrated a new concept for a mm-scale surface trap that incorporates good aspects of the two main trap categories; micro- and macroscopic. It can be situated outside the vacuum region since the trap centre is at \SI{2}{mm} from its surface and its power consumption is modest ($\sim$\SI{1}{kW}). Although there is scope for further improvements of this design in future iterations, it is capable of producing very high gradients ($>\SI{1000}{G/cm}$) and can be water-cooled very efficiently because of its small size and large surface area. Cold atoms applications where compactness is important could greatly benefit from this trap concept.

\begin{acknowledgments}
The authors would like to thank Graham Quelch for technical support. We acknowledge support from the Bodossaki foundation (DT), St. Peter's College and partial support from EPSRC.
\end{acknowledgments}

\bibliography{MesoscopicTrap.bib}

\begin{thebibliography}{19}%
\makeatletter
\providecommand \@ifxundefined [1]{%
 \@ifx{#1\undefined}
}%
\providecommand \@ifnum [1]{%
 \ifnum #1\expandafter \@firstoftwo
 \else \expandafter \@secondoftwo
 \fi
}%
\providecommand \@ifx [1]{%
 \ifx #1\expandafter \@firstoftwo
 \else \expandafter \@secondoftwo
 \fi
}%
\providecommand \natexlab [1]{#1}%
\providecommand \enquote  [1]{``#1''}%
\providecommand \bibnamefont  [1]{#1}%
\providecommand \bibfnamefont [1]{#1}%
\providecommand \citenamefont [1]{#1}%
\providecommand \href@noop [0]{\@secondoftwo}%
\providecommand \href [0]{\begingroup \@sanitize@url \@href}%
\providecommand \@href[1]{\@@startlink{#1}\@@href}%
\providecommand \@@href[1]{\endgroup#1\@@endlink}%
\providecommand \@sanitize@url [0]{\catcode `\\12\catcode `\$12\catcode
  `\&12\catcode `\#12\catcode `\^12\catcode `\_12\catcode `\%12\relax}%
\providecommand \@@startlink[1]{}%
\providecommand \@@endlink[0]{}%
\providecommand \url  [0]{\begingroup\@sanitize@url \@url }%
\providecommand \@url [1]{\endgroup\@href {#1}{\urlprefix }}%
\providecommand \urlprefix  [0]{URL }%
\providecommand \Eprint [0]{\href }%
\providecommand \doibase [0]{http://dx.doi.org/}%
\providecommand \selectlanguage [0]{\@gobble}%
\providecommand \bibinfo  [0]{\@secondoftwo}%
\providecommand \bibfield  [0]{\@secondoftwo}%
\providecommand \translation [1]{[#1]}%
\providecommand \BibitemOpen [0]{}%
\providecommand \bibitemStop [0]{}%
\providecommand \bibitemNoStop [0]{.\EOS\space}%
\providecommand \EOS [0]{\spacefactor3000\relax}%
\providecommand \BibitemShut  [1]{\csname bibitem#1\endcsname}%
\let\auto@bib@innerbib\@empty
\bibitem [{\citenamefont {Anderson}\ \emph {et~al.}(1995)\citenamefont
  {Anderson}, \citenamefont {Ensher}, \citenamefont {Matthews}, \citenamefont
  {Wieman},\ and\ \citenamefont {Cornell}}]{anderson_observation_1995}%
  \BibitemOpen
  \bibfield  {author} {\bibinfo {author} {\bibfnamefont {M.~H.}\ \bibnamefont
  {Anderson}}, \bibinfo {author} {\bibfnamefont {J.~R.}\ \bibnamefont
  {Ensher}}, \bibinfo {author} {\bibfnamefont {M.~R.}\ \bibnamefont
  {Matthews}}, \bibinfo {author} {\bibfnamefont {C.~E.}\ \bibnamefont
  {Wieman}}, \ and\ \bibinfo {author} {\bibfnamefont {E.~A.}\ \bibnamefont
  {Cornell}},\ }\href {\doibase 10.1126/science.269.5221.198} {\bibfield
  {journal} {\bibinfo  {journal} {Science}\ }\textbf {\bibinfo {volume}
  {269}},\ \bibinfo {pages} {198} (\bibinfo {year} {1995})}\BibitemShut
  {NoStop}%
\bibitem [{\citenamefont {Mewes}\ \emph {et~al.}(1996)\citenamefont {Mewes},
  \citenamefont {Andrews}, \citenamefont {van Druten}, \citenamefont {Kurn},
  \citenamefont {Durfee},\ and\ \citenamefont
  {Ketterle}}]{mewes_bose-einstein_1996}%
  \BibitemOpen
  \bibfield  {author} {\bibinfo {author} {\bibfnamefont {M.-O.}\ \bibnamefont
  {Mewes}}, \bibinfo {author} {\bibfnamefont {M.~R.}\ \bibnamefont {Andrews}},
  \bibinfo {author} {\bibfnamefont {N.~J.}\ \bibnamefont {van Druten}},
  \bibinfo {author} {\bibfnamefont {D.~M.}\ \bibnamefont {Kurn}}, \bibinfo
  {author} {\bibfnamefont {D.~S.}\ \bibnamefont {Durfee}}, \ and\ \bibinfo
  {author} {\bibfnamefont {W.}~\bibnamefont {Ketterle}},\ }\href {\doibase
  10.1103/PhysRevLett.77.416} {\bibfield  {journal} {\bibinfo  {journal}
  {Physical Review Letters}\ }\textbf {\bibinfo {volume} {77}},\ \bibinfo
  {pages} {416} (\bibinfo {year} {1996})}\BibitemShut {NoStop}%
\bibitem [{\citenamefont {Davis}\ \emph {et~al.}(1995)\citenamefont {Davis},
  \citenamefont {Mewes}, \citenamefont {Andrews}, \citenamefont {van Druten},
  \citenamefont {Durfee}, \citenamefont {Kurn},\ and\ \citenamefont
  {Ketterle}}]{davis_bose-einstein_1995}%
  \BibitemOpen
  \bibfield  {author} {\bibinfo {author} {\bibfnamefont {K.~B.}\ \bibnamefont
  {Davis}}, \bibinfo {author} {\bibfnamefont {M.~O.}\ \bibnamefont {Mewes}},
  \bibinfo {author} {\bibfnamefont {M.~R.}\ \bibnamefont {Andrews}}, \bibinfo
  {author} {\bibfnamefont {N.~J.}\ \bibnamefont {van Druten}}, \bibinfo
  {author} {\bibfnamefont {D.~S.}\ \bibnamefont {Durfee}}, \bibinfo {author}
  {\bibfnamefont {D.~M.}\ \bibnamefont {Kurn}}, \ and\ \bibinfo {author}
  {\bibfnamefont {W.}~\bibnamefont {Ketterle}},\ }\href {\doibase
  10.1103/PhysRevLett.75.3969} {\bibfield  {journal} {\bibinfo  {journal}
  {Physical Review Letters}\ }\textbf {\bibinfo {volume} {75}},\ \bibinfo
  {pages} {3969} (\bibinfo {year} {1995})}\BibitemShut {NoStop}%
\bibitem [{\citenamefont {Gildemeister}\ \emph {et~al.}(2010)\citenamefont
  {Gildemeister}, \citenamefont {Nugent}, \citenamefont {Sherlock},
  \citenamefont {Kubasik}, \citenamefont {Sheard},\ and\ \citenamefont
  {Foot}}]{gildemeister_trapping_2010}%
  \BibitemOpen
  \bibfield  {author} {\bibinfo {author} {\bibfnamefont {M.}~\bibnamefont
  {Gildemeister}}, \bibinfo {author} {\bibfnamefont {E.}~\bibnamefont
  {Nugent}}, \bibinfo {author} {\bibfnamefont {B.~E.}\ \bibnamefont
  {Sherlock}}, \bibinfo {author} {\bibfnamefont {M.}~\bibnamefont {Kubasik}},
  \bibinfo {author} {\bibfnamefont {B.~T.}\ \bibnamefont {Sheard}}, \ and\
  \bibinfo {author} {\bibfnamefont {C.~J.}\ \bibnamefont {Foot}},\ }\href
  {\doibase 10.1103/PhysRevA.81.031402} {\bibfield  {journal} {\bibinfo
  {journal} {Physical Review A}\ }\textbf {\bibinfo {volume} {81}},\ \bibinfo
  {pages} {031402} (\bibinfo {year} {2010})}\BibitemShut {NoStop}%
\bibitem [{\citenamefont {Fortágh}\ and\ \citenamefont
  {Zimmermann}(2007)}]{fortagh_magnetic_2007}%
  \BibitemOpen
  \bibfield  {author} {\bibinfo {author} {\bibfnamefont {J.}~\bibnamefont
  {Fortágh}}\ and\ \bibinfo {author} {\bibfnamefont {C.}~\bibnamefont
  {Zimmermann}},\ }\href {\doibase 10.1103/RevModPhys.79.235} {\bibfield
  {journal} {\bibinfo  {journal} {Reviews of Modern Physics}\ }\textbf
  {\bibinfo {volume} {79}},\ \bibinfo {pages} {235} (\bibinfo {year}
  {2007})}\BibitemShut {NoStop}%
\bibitem [{\citenamefont {Reichel}\ \emph {et~al.}(1999)\citenamefont
  {Reichel}, \citenamefont {Hänsel},\ and\ \citenamefont
  {Hänsch}}]{reichel_atomic_1999}%
  \BibitemOpen
  \bibfield  {author} {\bibinfo {author} {\bibfnamefont {J.}~\bibnamefont
  {Reichel}}, \bibinfo {author} {\bibfnamefont {W.}~\bibnamefont {Hänsel}}, \
  and\ \bibinfo {author} {\bibfnamefont {T.~W.}\ \bibnamefont {Hänsch}},\
  }\href {\doibase 10.1103/PhysRevLett.83.3398} {\bibfield  {journal} {\bibinfo
   {journal} {Physical Review Letters}\ }\textbf {\bibinfo {volume} {83}},\
  \bibinfo {pages} {3398} (\bibinfo {year} {1999})}\BibitemShut {NoStop}%
\bibitem [{\citenamefont {Hinds}\ and\ \citenamefont
  {Hughes}(1999)}]{hinds_magnetic_1999}%
  \BibitemOpen
  \bibfield  {author} {\bibinfo {author} {\bibfnamefont {E.~A.}\ \bibnamefont
  {Hinds}}\ and\ \bibinfo {author} {\bibfnamefont {I.~G.}\ \bibnamefont
  {Hughes}},\ }\href {\doibase 10.1088/0022-3727/32/18/201} {\bibfield
  {journal} {\bibinfo  {journal} {Journal of Physics D: Applied Physics}\
  }\textbf {\bibinfo {volume} {32}},\ \bibinfo {pages} {R119} (\bibinfo {year}
  {1999})}\BibitemShut {NoStop}%
\bibitem [{\citenamefont {Fortagh}\ \emph {et~al.}(1998)\citenamefont
  {Fortagh}, \citenamefont {Grossmann}, \citenamefont {Zimmermann},\ and\
  \citenamefont {Hänsch}}]{fortagh_miniaturized_1998}%
  \BibitemOpen
  \bibfield  {author} {\bibinfo {author} {\bibfnamefont {J.}~\bibnamefont
  {Fortagh}}, \bibinfo {author} {\bibfnamefont {A.}~\bibnamefont {Grossmann}},
  \bibinfo {author} {\bibfnamefont {C.}~\bibnamefont {Zimmermann}}, \ and\
  \bibinfo {author} {\bibfnamefont {T.~W.}\ \bibnamefont {Hänsch}},\ }\href
  {\doibase 10.1103/PhysRevLett.81.5310} {\bibfield  {journal} {\bibinfo
  {journal} {Physical Review Letters}\ }\textbf {\bibinfo {volume} {81}},\
  \bibinfo {pages} {5310} (\bibinfo {year} {1998})}\BibitemShut {NoStop}%
\bibitem [{\citenamefont {Schmiedmayer}(1998)}]{schmiedmayer_quantum_1998}%
  \BibitemOpen
  \bibfield  {author} {\bibinfo {author} {\bibfnamefont {J.}~\bibnamefont
  {Schmiedmayer}},\ }\href {\doibase 10.1007/s100530050184} {\bibfield
  {journal} {\bibinfo  {journal} {The European Physical Journal D - Atomic,
  Molecular, Optical and Plasma Physics}\ }\textbf {\bibinfo {volume} {4}},\
  \bibinfo {pages} {57} (\bibinfo {year} {1998})}\BibitemShut {NoStop}%
\bibitem [{\citenamefont {Krüger}\ \emph {et~al.}(2003)\citenamefont
  {Krüger}, \citenamefont {Luo}, \citenamefont {Klein}, \citenamefont
  {Brugger}, \citenamefont {Haase}, \citenamefont {Wildermuth}, \citenamefont
  {Groth}, \citenamefont {Bar-Joseph}, \citenamefont {Folman},\ and\
  \citenamefont {Schmiedmayer}}]{kruger_trapping_2003}%
  \BibitemOpen
  \bibfield  {author} {\bibinfo {author} {\bibfnamefont {P.}~\bibnamefont
  {Krüger}}, \bibinfo {author} {\bibfnamefont {X.}~\bibnamefont {Luo}},
  \bibinfo {author} {\bibfnamefont {M.~W.}\ \bibnamefont {Klein}}, \bibinfo
  {author} {\bibfnamefont {K.}~\bibnamefont {Brugger}}, \bibinfo {author}
  {\bibfnamefont {A.}~\bibnamefont {Haase}}, \bibinfo {author} {\bibfnamefont
  {S.}~\bibnamefont {Wildermuth}}, \bibinfo {author} {\bibfnamefont
  {S.}~\bibnamefont {Groth}}, \bibinfo {author} {\bibfnamefont
  {I.}~\bibnamefont {Bar-Joseph}}, \bibinfo {author} {\bibfnamefont
  {R.}~\bibnamefont {Folman}}, \ and\ \bibinfo {author} {\bibfnamefont
  {J.}~\bibnamefont {Schmiedmayer}},\ }\href {\doibase
  10.1103/PhysRevLett.91.233201} {\bibfield  {journal} {\bibinfo  {journal}
  {Physical Review Letters}\ }\textbf {\bibinfo {volume} {91}},\ \bibinfo
  {pages} {233201} (\bibinfo {year} {2003})}\BibitemShut {NoStop}%
\bibitem [{\citenamefont {Weinstein}\ and\ \citenamefont
  {Libbrecht}(1995)}]{weinstein_microscopic_1995}%
  \BibitemOpen
  \bibfield  {author} {\bibinfo {author} {\bibfnamefont {J.~D.}\ \bibnamefont
  {Weinstein}}\ and\ \bibinfo {author} {\bibfnamefont {K.~G.}\ \bibnamefont
  {Libbrecht}},\ }\href {\doibase 10.1103/PhysRevA.52.4004} {\bibfield
  {journal} {\bibinfo  {journal} {Physical Review A}\ }\textbf {\bibinfo
  {volume} {52}},\ \bibinfo {pages} {4004} (\bibinfo {year}
  {1995})}\BibitemShut {NoStop}%
\bibitem [{\citenamefont {Gupta}\ \emph {et~al.}(2005)\citenamefont {Gupta},
  \citenamefont {Murch}, \citenamefont {Moore}, \citenamefont {Purdy},\ and\
  \citenamefont {Stamper-Kurn}}]{gupta_bose-einstein_2005}%
  \BibitemOpen
  \bibfield  {author} {\bibinfo {author} {\bibfnamefont {S.}~\bibnamefont
  {Gupta}}, \bibinfo {author} {\bibfnamefont {K.~W.}\ \bibnamefont {Murch}},
  \bibinfo {author} {\bibfnamefont {K.~L.}\ \bibnamefont {Moore}}, \bibinfo
  {author} {\bibfnamefont {T.~P.}\ \bibnamefont {Purdy}}, \ and\ \bibinfo
  {author} {\bibfnamefont {D.~M.}\ \bibnamefont {Stamper-Kurn}},\ }\href
  {\doibase 10.1103/PhysRevLett.95.143201} {\bibfield  {journal} {\bibinfo
  {journal} {Physical Review Letters}\ }\textbf {\bibinfo {volume} {95}},\
  \bibinfo {pages} {143201} (\bibinfo {year} {2005})}\BibitemShut {NoStop}%
\bibitem [{\citenamefont {Jöllenbeck}\ \emph {et~al.}(2011)\citenamefont
  {Jöllenbeck}, \citenamefont {Mahnke}, \citenamefont {Randoll}, \citenamefont
  {Ertmer}, \citenamefont {Arlt},\ and\ \citenamefont
  {Klempt}}]{jollenbeck_hexapole-compensated_2011}%
  \BibitemOpen
  \bibfield  {author} {\bibinfo {author} {\bibfnamefont {S.}~\bibnamefont
  {Jöllenbeck}}, \bibinfo {author} {\bibfnamefont {J.}~\bibnamefont {Mahnke}},
  \bibinfo {author} {\bibfnamefont {R.}~\bibnamefont {Randoll}}, \bibinfo
  {author} {\bibfnamefont {W.}~\bibnamefont {Ertmer}}, \bibinfo {author}
  {\bibfnamefont {J.}~\bibnamefont {Arlt}}, \ and\ \bibinfo {author}
  {\bibfnamefont {C.}~\bibnamefont {Klempt}},\ }\href {\doibase
  10.1103/PhysRevA.83.043406} {\bibfield  {journal} {\bibinfo  {journal}
  {Physical Review A}\ }\textbf {\bibinfo {volume} {83}},\ \bibinfo {pages}
  {043406} (\bibinfo {year} {2011})}\BibitemShut {NoStop}%
\bibitem [{\citenamefont {Hopkins}\ \emph {et~al.}(2004)\citenamefont
  {Hopkins}, \citenamefont {Lev},\ and\ \citenamefont
  {Mabuchi}}]{hopkins_proposed_2004}%
  \BibitemOpen
  \bibfield  {author} {\bibinfo {author} {\bibfnamefont {A.}~\bibnamefont
  {Hopkins}}, \bibinfo {author} {\bibfnamefont {B.}~\bibnamefont {Lev}}, \ and\
  \bibinfo {author} {\bibfnamefont {H.}~\bibnamefont {Mabuchi}},\ }\href
  {\doibase 10.1103/PhysRevA.70.053616} {\bibfield  {journal} {\bibinfo
  {journal} {Physical Review A}\ }\textbf {\bibinfo {volume} {70}},\ \bibinfo
  {pages} {053616} (\bibinfo {year} {2004})}\BibitemShut {NoStop}%
\bibitem [{\citenamefont {Wang}\ \emph {et~al.}(2007)\citenamefont {Wang},
  \citenamefont {Liu}, \citenamefont {Minardi},\ and\ \citenamefont
  {Kasevich}}]{wang_reaching_2007}%
  \BibitemOpen
  \bibfield  {author} {\bibinfo {author} {\bibfnamefont {R.}~\bibnamefont
  {Wang}}, \bibinfo {author} {\bibfnamefont {M.}~\bibnamefont {Liu}}, \bibinfo
  {author} {\bibfnamefont {F.}~\bibnamefont {Minardi}}, \ and\ \bibinfo
  {author} {\bibfnamefont {M.}~\bibnamefont {Kasevich}},\ }\href {\doibase
  10.1103/PhysRevA.75.013610} {\bibfield  {journal} {\bibinfo  {journal}
  {Physical Review A}\ }\textbf {\bibinfo {volume} {75}},\ \bibinfo {pages}
  {013610} (\bibinfo {year} {2007})}\BibitemShut {NoStop}%
\bibitem [{\citenamefont {Jian}\ and\ \citenamefont {van
  Wijngaarden}(2013)}]{jian_double-loop_2013}%
  \BibitemOpen
  \bibfield  {author} {\bibinfo {author} {\bibfnamefont {B.}~\bibnamefont
  {Jian}}\ and\ \bibinfo {author} {\bibfnamefont {W.~A.}\ \bibnamefont {van
  Wijngaarden}},\ }\href {\doibase 10.1364/JOSAB.30.000238} {\bibfield
  {journal} {\bibinfo  {journal} {Journal of the Optical Society of America B}\
  }\textbf {\bibinfo {volume} {30}},\ \bibinfo {pages} {238} (\bibinfo {year}
  {2013})}\BibitemShut {NoStop}%
\bibitem [{\citenamefont {Chubar}\ \emph {et~al.}(1998)\citenamefont {Chubar},
  \citenamefont {Elleaume},\ and\ \citenamefont
  {Chavanne}}]{chubar_three-dimensional_1998}%
  \BibitemOpen
  \bibfield  {author} {\bibinfo {author} {\bibfnamefont {O.}~\bibnamefont
  {Chubar}}, \bibinfo {author} {\bibfnamefont {P.}~\bibnamefont {Elleaume}}, \
  and\ \bibinfo {author} {\bibfnamefont {J.}~\bibnamefont {Chavanne}},\ }\href
  {\doibase 10.1107/S0909049597013502} {\bibfield  {journal} {\bibinfo
  {journal} {Journal of Synchrotron Radiation}\ }\textbf {\bibinfo {volume}
  {5}},\ \bibinfo {pages} {481} (\bibinfo {year} {1998})}\BibitemShut {NoStop}%
\bibitem [{\citenamefont {Sukumar}\ and\ \citenamefont
  {Brink}(1997)}]{sukumar_spin-flip_1997}%
  \BibitemOpen
  \bibfield  {author} {\bibinfo {author} {\bibfnamefont {C.~V.}\ \bibnamefont
  {Sukumar}}\ and\ \bibinfo {author} {\bibfnamefont {D.~M.}\ \bibnamefont
  {Brink}},\ }\href {\doibase 10.1103/PhysRevA.56.2451} {\bibfield  {journal}
  {\bibinfo  {journal} {Physical Review A}\ }\textbf {\bibinfo {volume} {56}},\
  \bibinfo {pages} {2451} (\bibinfo {year} {1997})}\BibitemShut {NoStop}%
\bibitem [{Note1()}]{Note1}%
  \BibitemOpen
  \bibinfo {note} {The Computer Aided Engineering Linux distribution \protect
  \href {http://www.caelinux.com/CMS/}{http://www.caelinux.com/CMS/} is based
  on Ubuntu and includes a plethora of open-source software. The SALOME project
  \protect \href
  {http://www.salome-platform.org/}{http://www.salome-platform.org/} aims to
  develop an open-source CAD-to-FEA link based on Open CASCADE. Elmer \protect
  \href
  {http://www.csc.fi/english/pages/elmer}{http://www.csc.fi/english/pages/elmer}
  is an open-source multiphysical simulation software that uses the Finite
  Element Method to solve the partial differential equations describing a
  physical system.}\BibitemShut {Stop}%
\end{thebibliography}%

\end{document}